\documentclass[12pt]{iopart}
\usepackage{epsfig}
\begin{document}

\title{Secure communication with single-photon two-qubit states}

\author{Almut Beige \dag, Berthold-Georg Englert \dag \ddag, \\
Christian Kurtsiefer \S, and Harald Weinfurter \dag \S }

\address{\dag Max-Planck-Institut f\"ur Quantenoptik, Hans-Kopfermann-Str. 1,  
85748 Garching, Germany}
\address{\ddag Atominstitut, Technische Universit\"at Wien, Stadionallee 2, 1020 
Wien, Austria}
\address{\S Sektion Physik, Universit\"at M\"unchen, Schellingstrasse 4, 80799 
M\"unchen, Germany}

\begin{abstract}
We propose a cryptographic scheme that is deterministic:
Alice sends single photons to Bob, and each and every photon detected supplies
one key bit --- no photon is wasted. 
This is in marked contrast to other schemes in which a random process decides
whether the next photon sent will contribute to the key or not. 
The determinism is achieved by preparing the photons in two-qubit states,
rather than the one-qubit states used in conventional schemes.
In particular, we consider the realistic situation in which one qubit is the
photon polarization, the other a spatial alternative. 
Further, we show how one can exploit the deterministic nature
for direct secure communication,
that is: \emph{without} the need for establishing a shared key first. 
\end{abstract}


\submitto{\JPA}

\maketitle

\section{Introduction}

Cryptographic schemes based on the exchange of single photons, each 
carrying one bit of information, have been widely discussed in the 
literature \cite{reviews}. 
In some of the schemes, \emph{Alice} and \emph{Bob} share  
entangled photon pairs \cite{Ekert}. In others, Bob performs 
measurements on photons that Alice sends him \cite{Bennett}.
They always need to communicate via a classical channel as well. 
Experiments have shown that secure key distribution is possible
indeed, even over a distance of several kilometers \cite{Hughes,Gisin,recent}.

The standard procedures, such as the so-called BB84 protocol of
\cite{Bennett}, are \emph{not} deterministic in the sense that Bob may or may
not get a key bit for the next photon that Alice will send;
on average one key bit is obtained for every two photons 
transmitted\footnote{Only few protocols try to go beyond this 50\% 
efficiency \cite{LoCab}.}. 
By contrast, the scheme we propose here, \emph{is} deterministic:
Bob gets a key bit for each and every photon sent by Alice.

This determinism is the main advantage of our new scheme.
It offers, in particular, the option of secure communication without first
establishing a shared key.

To achieve the determinism,
Alice sends Bob photons prepared in certain two-qubit states,
rather than photons carrying one-qubit states.  
She uses, for example,
the spatial binary alternative of the photon with the basis states 
$|{\sf R}\rangle$ and  $|{\sf L}\rangle$ and the two polarization states 
$|{\sf v}\rangle$ and $|{\sf h}\rangle$. 
Here, $|{\sf R}\rangle$ and $|{\sf L}\rangle$ describe
a photon traveling in the ``right'' or the ``left'' fiber, respectively, 
and $|{\sf v}\rangle$ and $|{\sf h}\rangle$ refer to photons with 
vertical and horizontal polarization. 

With the aid of unitary two-qubit gates \cite{Englert}, 
Alice can turn either one of the simple product states 
$|{\sf Rv}\rangle=|{\sf R}\rangle\otimes|{\sf v}\rangle$, $|{\sf Rh}\rangle$, 
$|{\sf Lv}\rangle$, and $|{\sf Lh}\rangle$ into any desired superposition 
thereof, so that she can send each photon in the single-photon two-qubit 
state of her choosing. 
Likewise, Bob's measurements of certain sets of four mutually 
orthogonal two-qubit states are achieved by appropriate unitary gates. 
They transform the states of the measurement basis in question into the 
four basic product states, which are then easily discriminated.

Since each photon carries two qubits now, the new scheme 
is not more efficient in terms of qubits than the standard ones:
Each qubit pair sent gives one key bit.

\section{Quantum key distribution}

We present the scheme for key distribution first. 
It has, of course, a number of features in common with the BB84 protocol,
but generates a key bit for every transmitted photon.
We discuss its security against eavesdropping and observe that, despite the
deterministic nature, it cannot be used for \emph{direct} communication, 
that is: for sending a message without establishing a shared key first.
We then introduce a second scheme,
with more involved state preparation and analysis, 
that does enable Alice and Bob to communicate directly and confidentially.

\begin{figure}
\hspace*{2.4cm} \epsfig{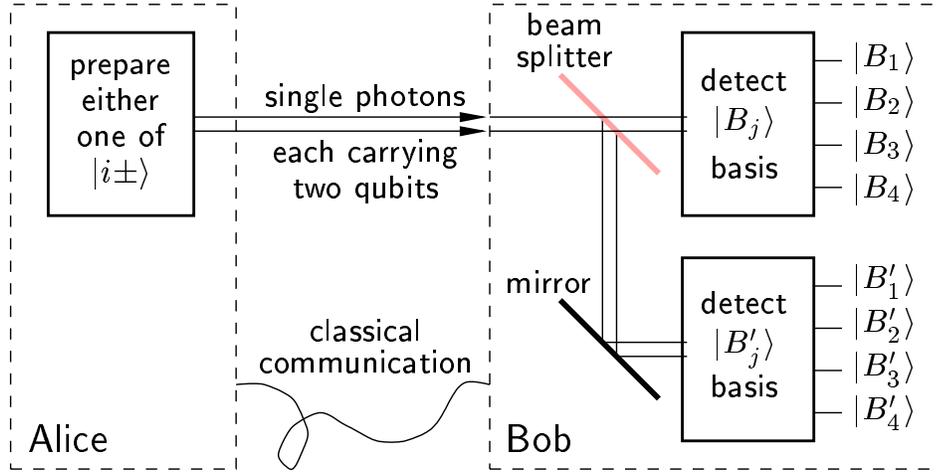}
\caption[Setup]{\label{fig:setup}%
Schematic view of the setup.
To transmit the bit ``$+$'' or ``$-$'' Alice  
prepares a photon in one of the states $|i\pm \rangle$ and
sends it to Bob. 
For the purpose of cryptography, two pairs of states are sufficient 
[e.g., those of (\ref{eq:k-states})], whereas communication requires 
four pairs [those of (\ref{eq:3-1pairs})].
A beam splitter reroutes the photon randomly, 
and then Bob measures in which state $|B_j\rangle$ or 
$|B'_j\rangle$ it is. 
In some cases, communication via the classical 
channel is required to decode the bit.} 
\end{figure}

The experimental setup of the cryptographic scheme we propose is sketched 
in Fig.~\ref{fig:setup}. To transmit one bit, 
``$+$'' or ``$-$'', Alice sends Bob a photon prepared in one
of the four states $|i \pm \rangle $ $(i=1,2)$.
For a ``$+$'' bit she chooses randomly between $|1+\rangle$ and $|2+\rangle$;
for a ``$-$'' bit between $|1- \rangle$ and $|2-\rangle$.
When the photon arrives at Bob's end, he randomly chooses between two 
different two-qubit bases for his analysis of the photon state. 
Experimentally, this can be achieved by sending the incoming 
photon through a beam splitter and rerouting it to different measurement 
devices as shown in Fig.~\ref{fig:setup}. 
Bob measures either the basis states $|B_1 \rangle,\,\dots,\,|B_4 \rangle$ 
or $|B'_1\rangle,\,\dots,\, |B'_4 \rangle$,
and he can \emph{always} infer the bit Alice sent.
Depending on the outcome of his measurement, he might be able to deduce 
the incoming bit immediately. 
In some cases, classical communication is required, and Alice has to tell 
Bob whether the photon state she prepared was of type ``$1$'' or ``$2$''. 

For illustration, let us consider the simplest version in which the states
sent by Alice and the states detected by Bob are all product states\footnote{Therefore, 
this particular example could also be realized by exploiting the
polarization qubits of paired photons.}
such as
\begin{eqnarray}
\bigl(|1+\rangle,|1-\rangle;|2+\rangle,|2-\rangle\bigr)
&=&\bigl(|{\sf Rs}\rangle,|{\sf La}\rangle;
|{\sf Sv}\rangle,|{\sf Ah}\rangle\bigr)\,,
\nonumber\\
\bigl(|B_1\rangle,|B_2\rangle,|B_3\rangle,|B_4\rangle\bigr)
&=&
\bigl(|{\sf Rv}\rangle,|{\sf Rh}\rangle,|{\sf Lv}\rangle,|{\sf Lh}\rangle\bigr)
\,,\nonumber\\
\bigl(|B'_1\rangle,|B'_2\rangle,|B'_3\rangle,|B'_4\rangle\bigr)
&=&
\bigl(|{\sf Ss}\rangle,|{\sf As}\rangle,|{\sf Sa}\rangle,|{\sf Aa}\rangle\bigr)
  \label{eq:products}
\end{eqnarray}
where
\begin{equation}
  \label{eq:sym-asym}
\left.\begin{array}{l} |{\sf S}\rangle \\ |{\sf A}\rangle \end{array}\right\}
=\frac{1}{\sqrt{2}}\bigl(|{\sf R}\rangle\pm|{\sf L}\rangle\bigr)
\,,\quad
\left.\begin{array}{l} |{\sf s}\rangle \\ |{\sf a}\rangle \end{array}\right\}
=\frac{1}{\sqrt{2}}\bigl(|{\sf v}\rangle\pm|{\sf h}\rangle\bigr)
\end{equation}
are symmetric (${\sf S}$ and ${\sf s}$) and antisymmetric 
(${\sf A}$ and ${\sf a}$) superpositions of the basic alternatives.
Note that each of Bob's states is orthogonal to either the ``$+$'' state
or the ``$-$'' state of each pair; this is the essential property for the
deterministic transmission.
Suppose, for instance, that Bob detects state $|B_3\rangle=|{\sf Lv}\rangle$;
it is orthogonal to ${|1+\rangle=|{\sf Rs}\rangle}$ and 
${|2-\rangle=|{\sf Ah}\rangle}$ and therefore it signifies ``$-$'' if a photon
of type ``$1$'' was sent and ``$+$'' if it was of type ``$2$''.
These matters are summarized in Table \ref{tbl:keybits}.

\begin{table}
\caption{\label{tbl:keybits}%
For the states of (\ref{eq:products}):
Key bits as inferred by Bob upon learning which type of photon was sent by
Alice.
Note that Bob does not need this classical information if he detects the
states of the 1st and 4th columns. }
\begin{indented}
\item[] \begin{tabular}{c|cccc}
\br 
photon sent & \multicolumn{4}{c}{state detected by Bob}\\
by Alice & $B_1$ or $B'_1$ & $B_2$ or $B'_2$& $B_3$ or $B'_3$&
$B_4$ or $B'_4$ \\ 
\mr
type $1$ & $+$ & $+$ & $-$ & $-$ \\
type $2$ & $+$ & $-$ & $+$ & $-$ \\
\br
\end{tabular}
\end{indented}
\end{table}

Let us now exhibit the basic general features of our deterministic scheme, as
they are illustrated by this particular example.
How can Bob always know which bit Alice sent?
He can distinguish the ``$+$'' states from the ``$-$'' states unambiguously
if, for all state pairs ${|i+\rangle/|i-\rangle}$, each possible measurement
result can only be caused by $|i+\rangle$ or $|i-\rangle$, but not by both.
This must be the case for every basis measured by Bob.
Then he can infer the bit transmitted as soon as Alice identifies the type of
pair used (that is: she tells him the value of the pair label $i$).

For the security of the scheme it is important that the state pairs
$|1\pm\rangle$ and $|2\pm\rangle$ sent by Alice are neither identical nor
orthogonal. 
It is equally important that Bob has more than one basis at his disposal, 
because this is what renders possible the detection of an eavesdropper.
In the example (\ref{eq:products}), the two bases are in fact even
complementary since the transition probabilities 
${\bigl|\langle B_i|B'_j\rangle\bigr|^2=\frac{1}{4}}$ do not depend on the 
quantum numbers $i,j$.
This maximal incompatibility is not really needed, but the bases should not be
very similar to each other in order to ensure that an eavesdropper will  
surely cause a substantial number of false detections at Bob's end, 
as we discuss below.

To analyze this in more detail, we consider a scheme that is somewhat more
general than the one based on the products states (\ref{eq:products}).
Here Bob's bases are related to each other by
\begin{equation} \label{eq:B->B'}
\bigl(|B'_1\rangle, |B'_2\rangle, |B'_3\rangle, |B'_4\rangle \bigr) 
= \bigl(|B_1\rangle, |B_2\rangle, |B_3\rangle, 
|B_4\rangle \bigr){\bf K}  
\end{equation}
where the $4\times4$ matrix ${\bf K}$ is given by 
\begin{equation} \label{eq:4x4K}
{\bf K} = \frac{1}{1+k^2}
\left( \begin{array}{llll}
1 & \phantom{-}k & \phantom{-}k & \phantom{-}k^2 \\
k & \phantom{-}k^2 & -1 & -k \\
k & -1 & \phantom{-}k^2 & -k \\
k^2 & -k & -k & \phantom{-}1 \end{array} \right)
\end{equation}
with a real parameter $k$.
For brevity and simplicity, we are satisfied with discussing the most
elementary version of the scheme, where Alice makes use of two state pairs 
only that are given by
\begin{eqnarray} 
|1+\rangle &=& \bigl(|B_1\rangle + k |B_2 \rangle\bigr)/\sqrt{1+k^2}\,, 
\nonumber \\
|1-\rangle &=& \bigl(k|B_3\rangle-|B_4\rangle\bigr)/\sqrt{1+k^2}\,, 
\nonumber \\
|2+\rangle &=& \bigl(|B_1\rangle+k|B_3\rangle\bigr)/\sqrt{1+k^2}\,, 
\nonumber \\
|2-\rangle &=& \bigl(k|B_2\rangle-|B_4\rangle\bigr)/\sqrt{1+k^2}\,, 
\label{eq:k-states}
\end{eqnarray}
More generally, she could always use four pairs, and even six pairs 
in some versions \cite{ToCome}.
Relations (\ref{eq:k-states}) remain valid if the $|B_j\rangle$'s 
are replaced by the $|B'_j\rangle$'s. 
Note that the inverse of the transformation (\ref{eq:B->B'}) is also furnished 
by ${\bf K}$ since this matrix is both Hermitian and unitary.
For ${k=1}$, in particular, we return to the situation of (\ref{eq:products})
where the two bases $|B_j\rangle$ and $|B'_j\rangle$ are complementary.
Table \ref{tbl:keybits} continues to apply, irrespective of the value of $k$. 

Let us now imagine that \emph{Evan}, the eavesdropper, is listening in.
He intercepts each photon sent by Alice, performs a measurement on it, and
then forwards a replacement photon to Bob.
Evan will not be able to infer with certainty which two-qubit state is
carried by the intercepted photon, and so  he has to make an educated guess 
based on his measurement result.
Then he prepares the replacement photon accordingly, namely in the two-qubit 
state that has the best chance of avoiding wrong detector clicks at Bob's end.
If, for instance, Alice has sent a $|1+\rangle$ photon, then the detectors for
$|B_3\rangle$ and $|B_4\rangle$ as well as $|B'_3\rangle$ and $|B'_4\rangle$
would yield wrong clicks and reveal the interference of the eavesdropper.
Thus, Evan has to solve a two-fold problem: 
Which basis should he measure, and which states should be forwarded to Bob,
such that the probability for a wrong click is minimal?

These questions can be answered systematically \cite{ToCome}, also for more
general intercept-resend strategies, much like the corresponding studies
\cite{FGGNP97} for the BB84 protocol.  
(The generalizations do not offer a real advantage to Evan, however.)
In an optimal strategy then, the probability that Bob will detect a
wrong click is 
\begin{equation}
  \label{eq:MinErrRate2}
  p^{(2)}_{\rm min}=\frac{1}{2}-\frac{1}{2}\frac{\sqrt{1+k^4}}{1+k^2}\;.
\end{equation}
All other intercept-resend strategies that Evan might employ result in larger 
error rates. 
The largest value obtains for ${k=\pm1}$, namely 
$p^{(2)}_{\rm min}=(2-\sqrt{2})/4=14.6\%$; for $k=0$ and $k\to\pm\infty$ the
minimal error rate vanishes.
Both limiting cases are easily understood: for $k=\pm1$ Bob's measurement bases
are complementary and therefore maximally incompatible, and for $k=0$ or
$k\to\pm\infty$ they are essentially identical.
We note in passing that, if four pairs of states are used rather than just the
two pairs of (\ref{eq:k-states}), the minimal error rate increases to
\begin{equation}
  \label{eq:MinErrRate4}
  p^{(4)}_{\rm min}=\frac{1}{2}\frac{\min\bigl\{1,k^2\bigr\}}{1+k^2}\;,
\end{equation}
which can be as large as $25\%$, and an
eavesdropper's presence can then be noticed more easily. 

For the purposes of this letter, we continue to focus on the two-pair scheme
and assume that Alice and Bob have wisely chosen a $k$ value near $k=1$, say.
Suppose they want to establish a key of 1000 bits, and Alice sends 1100
photons in two-qubit states, randomly chosen from the four states of
(\ref{eq:k-states}). 
Bob detects all photons, then selects a random subset of 100 and tells Alice
in which states he found them. 
If some of Bob's measurement results are inconsistent with the states 
Alice sent, such as detecting $|B'_3\rangle$ for a $|1+\rangle$ photon, 
then Alice doesn't trust the transmission and they start all over.
If, however, Bob's results are all right, then Alice concludes that the
likelihood that Evan has listened in is less than 
$(1-p^{(2)}_{\rm min})^{100}=1.3\times10^{-7}$, which she and Bob have earlier
decided to be sufficiently small for the security level they'd like to have.
Alice then reveals the type of each photon, ``$1$'' or ``$2$'', and Bob infers
the bits sent with the aid of Table \ref{tbl:keybits}.
Thereafter they share a secure  1000-bit key string.
A confidential message of this length can then be exchanged.

\section{Secure communication without first establishing a shared key}

Given the deterministic nature of the scheme, one might wonder if Alice
couldn't send a message directly to Bob without first establishing a shared
cryptographic key.
That would require that Evan cannot infer the transmitted bits before Alice
and Bob become aware of his presence.
Now, Table \ref{tbl:keybits} tells us that Evan could acquire correct
knowledge of every second bit sent by just performing the same
measurements as Bob because knowledge of the photon type is not needed 
in the 1st and 4th columns.
In fact, Evan can improve his educated guesses by choosing his measurement 
more cleverly \cite{ToCome}, since the ``$+$'' states sent by Alice are
distributed differently over the two-qubit Hilbert space than the ``$-$''
states. 
For the example of (\ref{eq:k-states}), he can
systematically exploit the difference between the two-dimensional subspaces
spanned by the ``$+$'' states and the ``$-$'' states to achieve odds as large as
${\frac{1}{2}+\frac{1}{2}/\sqrt{1+k^2}}$ for guessing the bits right, which
exceeds 85\% for ${k=1}$.
Clearly, secure direct communication is not possible under these
circumstances.

But there is a modified scheme that does enable Alice and Bob to communicate
directly and confidentially.
Again we focus on the simplest version, in which Bob's measurement bases are
related to each other by
\begin{equation}
  \label{eq:3-1bases}
\left(\begin{array}{l}
\langle B_1| \\ \langle B_2| \\ \langle B_3| \\ \langle B_4| 
\end{array}\right)=\frac{{\rm i}}{\sqrt{3}}
\left(\begin{array}{rrrr}
0 & 1 & 1 & 1 \\
-1 & 0 & -1 & 1 \\
-1 & 1 & 0 & -1 \\
-1 & -1 & 1 & 0
\end{array}\right)
\left(\begin{array}{l}
\langle B'_1| \\ \langle B'_2| \\ \langle B'_3| \\ \langle B'_4| 
\end{array}\right)\;.
\end{equation}
Just like ${\bf K}$ of (\ref{eq:4x4K}), the $4\times4$ transformation matrix
appearing here is Hermitian and unitary, so that it also furnishes the inverse
transformation.
The states sent by Alice now are identical with Bob's basis states, grouped
into four pairs of orthogonal states in accordance with
\begin{equation}
  \label{eq:3-1pairs}
  |i+\rangle=|B_i\rangle\,,\quad |i-\rangle=|B'_i\rangle
\quad {\rm for}~ i=1,\, 2, \, 3, \, 4 \;.
\end{equation}
The basic features discussed above in the paragraphs between
(\ref{eq:sym-asym}) and (\ref{eq:B->B'}) are here present as well.

\begin{table}
\caption{\label{tbl:keybits'}%
For the states of (\ref{eq:3-1bases}) and (\ref{eq:3-1pairs}):
Key bits as inferred by Bob upon learning which type of photon was sent by
Alice.}
\begin{indented}
\item[] \begin{tabular}{c|ccccccccc}
\br
photon sent & \multicolumn{9}{c}{state detected by Bob}\\
by Alice & $B_1$ & $B_2$ & $B_3$ & $B_4$  & 
& $B'_1$ & $B'_2$ & $B'_3$ & $B'_4$ \\ 
\mr
type $1$ & $+$ & $-$ & $-$ & $-$ && $-$ & $+$ & $+$ & $+$ \\
type $2$ & $-$ & $+$ & $-$ & $-$ && $+$ & $-$ & $+$ & $+$ \\
type $3$ & $-$ & $-$ & $+$ & $-$ && $+$ & $+$ & $-$ & $+$ \\
type $4$ & $-$ & $-$ & $-$ & $+$ && $+$ & $+$ & $+$ & $-$ \\
\br
\end{tabular}
\end{indented}
\end{table}

How Bob infers the bits sent is summarized in Table~\ref{tbl:keybits'}.
Consider, for example, that he found a certain photon in
state $|B_3\rangle$. 
He'll infer that ``$+$'' was sent if Alice tells him that it was a
\mbox{type-$3$} 
photon because $|B_3\rangle$ is orthogonal to $|3-\rangle$, 
and that ``$-$'' was sent if it was of type $1$, $2$, or $4$ because
$|B_3\rangle$ is orthogonal to $|1+\rangle$, $|2+\rangle$, and $|4+\rangle$.

Now, the minimal error rate resulting from eavesdropping of the
intercept-resend kind is $\frac{1}{6}=16.7\%$ for (\ref{eq:3-1pairs}) with
(\ref{eq:3-1bases}), which is less than the 25\% of the four-pair key-sharing 
scheme to which (\ref{eq:MinErrRate4}) refers, 
but more than the 14.6\% of the two-pair version\footnote{These error rates 
refer to the situation in which Evan wishes to find out the value of each bit 
transmitted. Instead, he could settle for just a reasonable likelihood for 
guessing the bit value right and bargain for a reduced error rate in return.
A detailed discussion of compromises of this kind will be presented elsewhere.}. 
Thus, Evan's interference can be detected just as easily in the present scheme
of (\ref{eq:3-1pairs}) and (\ref{eq:3-1bases}) as in the previous one of
(\ref{eq:k-states}) and (\ref{eq:B->B'}) or of (\ref{eq:products}).
Therefore, the scheme defined by (\ref{eq:3-1bases}) and (\ref{eq:3-1pairs})
could be used for secure key distribution.

But this scheme is also well suited for direct communication, 
since the four ``$+$'' states span the whole two-qubit Hilbert space
uniformly, and the four ``$-$'' states do so as well.
Thus, Evan cannot distinguish ``$+$'' photons from ``$-$''
photons here without knowing the photon type. 
In particular, although the columns of Table \ref{tbl:keybits'} have 3:1
ratios of the signs, both kinds carry equal weight.
If, for example, $|B_3\rangle$ is detected then $|3+\rangle$ is as likely as
$|1-\rangle$, $|2-\rangle$, and $|4-\rangle$ together. 

\begin{table}
\caption{\label{tbl:DirComm}%
Direct confidential communication.
Alice chooses a random key sequence of $1,2,3,4$ (1st row)
and matches it with the bit sequence of the message (2nd row)
interspersed with randomly located control bits (boxed)
to determined the sequence of states to be sent (3rd row).
Bob obtains a sequence of detected states (4th row).
The control bits are used to test for the presence of an eavesdropper.
After Alice reveals the random sequence of the 1st row, Bob can then
reconstruct the message of the 2nd row.}
\begin{indented}
\item[] \begin{tabular}{r@{\quad}ccccccccc@{$\,\cdots\;$}}
\br
Alice's key & 1 & 3 & 4 & 4 & 1 & 2 & 1 & 3 & 3 \\
message & $+$ & \fbox{$+$} & $-$ & $-$ & $-$ & $+$ 
& \fbox{$-$} & $+$ & $-$  \\
states sent & $1+$  & $3+$   & $4-$   & $4-$  & $1-$  & $2+$  & $1-$ 
& $3+$ & $3-$ \\
Bob finds & $B_1$ & $B'_1$ & $B'_4$ & $B_2$ & $B_2$ & $B'_4$ & $B_4$
& $B_3$ & $B'_3$ \\
\br 
\end{tabular}
\end{indented}
\end{table}

Direct confidential communication is achieved as follows; see Table
\ref{tbl:DirComm}. 
Step one: Alice generates, at her end, a random sequence of $1,2,3,4$ that will
serve as the cryptographic key.
Only Alice knows this key.
Step two: She matches this sequence with the string of $+/-$ message bits,
interspersed with a fair number of control bits at random positions, and so
determines the two-qubit states to be sent to Bob.
Only Alice knows which bits are control bits and which are message bits.
Step three: Alice sends the photons in these states, and Bob detects them in
one of the states of his measurement bases.
Step four: Alice tells Bob which photons carried control bits, and he tells
her in which state he found them.
Step five: Alice verifies that Bob's findings are consistent with what she
sent. 
If no inconsistencies --- that is: errors --- are noticed, Alice concludes
that the transmission was secure and continues with step six; 
otherwise she repeats the procedure beginning with step one.
Step six: Alice reveals the key sequence of step one, and Bob reconstructs 
the message with the aid of Table~\ref{tbl:keybits'}.

This scheme for direct communication is secure because Alice does not reveal
her key sequence until she has convinced herself that Evan has not been
listening in. 
Without this classical information, Evan cannot infer a single bit of the
message.  
The only bits he might decode before his presence
is detected are the control bits which, however, are not part of the
confidential message. 

\section{Final remarks}

Experimental implementations of our schemes for key distribution and direct
communication can be realized with the aid of the universal two-qubit gates
that where introduced recently \cite{Englert}.
Concerning practical aspects, we remark that we gain a factor of two compared
to other cryptography schemes even for imperfect transmission
and detection.
Redundant encoding can overcome the losses in the communication scheme without
revealing information to the eavesdropper.
We'd also like to  note that, 
rather than using two binary alternatives of single photons, one could, 
of course, equally well exploit the states of any other
four-dimensional Hilbert space.  

In summary, we propose a new cryptographic scheme. 
Under ideal conditions, the scheme is deterministic: 
Alice and Bob get a key bit for each photon sent, 
whereas other schemes \cite{Ekert,Bennett} need at least two photons
and are not deterministic. 
A significant percentage of Bob's measurement results will be wrong if an 
eavesdropper intercepts the transmission, so that his presence can 
surely be noticed. 
In addition, we show how the encoding and deterministic decoding of qubits in
a four-dimensional Hilbert space allows direct communication, even without
first establishing a shared key.

\ack 
A.~B. and B.-G.~E. are grateful for the hospitality at the
Erwin-Schr\"odinger-Institut in Vienna where part of this work was done.
Ch.~K. and H.~W.  acknowledge support 
by project QuComm (IST-1999-10033) of the European Union.

\end{document}